\documentclass{article}
\usepackage{spconf,amsmath,graphicx,hyperref,amssymb,booktabs,lipsum}
\usepackage[capitalize,noabbrev]{cleveref}
\usepackage{algorithm}
\usepackage{algpseudocode}
\usepackage{cite}

\title{\MakeUppercase{Scaling Multi-Talker ASR with Speaker-Agnostic Activity Streams}}
\name{
\begin{tabular}{c}
Xiluo He$^1$\thanks{This work was done at JSALT 2025. \\ \indent \indent Corresponding author: xhe69@jh.edu}, Alexander Polok$^2$, Jes\'us Villalba$^3$, Thomas Thebaud$^3$, Matthew Maciejewski$^3$
\end{tabular}}
\address{$^1$Department of Computer Science, Johns Hopkins University, Baltimore, USA \\
$^2$Faculty of Information Technology, Brno University of Technology, Brno, Czech Republic \\
$^3$Human Language Technology Center of Excellence, Johns Hopkins University, Baltimore, USA}
\begin{document}
%
\maketitle
\begin{abstract}
An increasingly common training paradigm for multi-talker automatic speech recognition (ASR) is to use speaker activity signals to adapt single-speaker ASR models for overlapping speech. Although effective, these systems require running the ASR model once per speaker, resulting in inference costs that scale with the number of speakers and limiting their practicality. In this work, we propose a method that decouples the inference cost of activity-conditioned ASR systems from the number of speakers by converting speaker-specific activity outputs into two speaker-agnostic streams. A central challenge is that naïvely merging speaker activities into streams significantly degrades recognition, since pretrained ASR models assume contiguous, single-speaker inputs. To address this, we design new heuristics aimed at preserving conversational continuity and maintaining compatibility with existing systems. We show that our approach is compatible with Diarization-Conditioned Whisper (DiCoW) to greatly reduce runtimes on the AMI and ICSI meeting datasets while retaining competitive performance. 
\end{abstract}
\section{Introduction}
\label{sec:intro}

Advances in deep learning and large-scale training have substantially improved automatic speech recognition (ASR) across diverse benchmarks \cite{radford2023robust, prabhavalkar2023end, li2022recent}. Building on these developments, multi-talker ASR systems have been proposed to extend recognition capabilities to overlapping, conversational speech. Despite significant progress, this domain remains highly challenging, as specialized modeling techniques are required to effectively address simultaneous speech and frequent turn-taking  \cite{yu2022comparative,cornell2024chime, abramovski2025-chime9nsfsummary}. 

\begin{figure}[t]
  \centering
  \centerline{\includegraphics[width=\columnwidth]{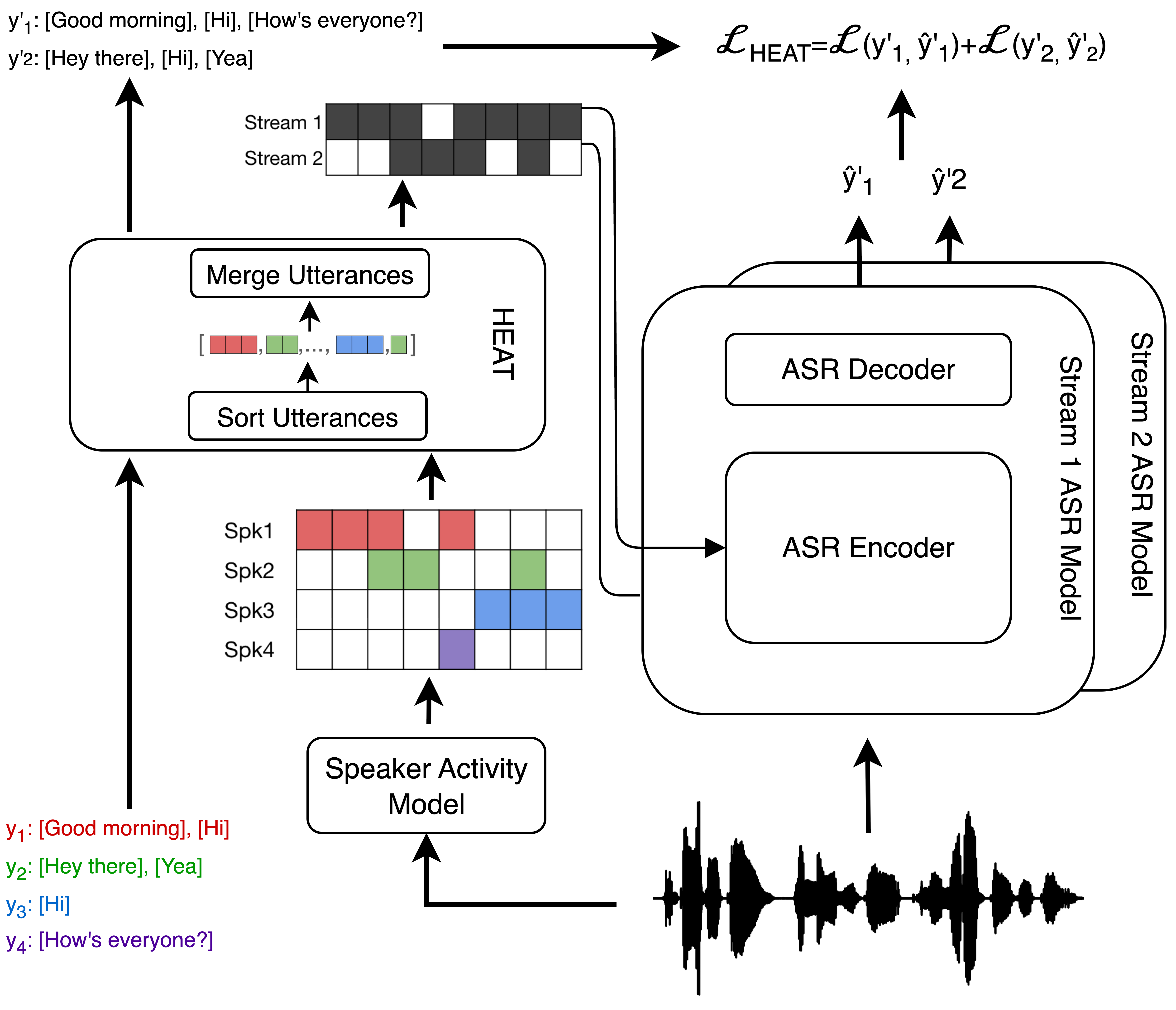}}
\caption{\textbf{Activity Conditioning with HEAT}. The ASR model runs twice ---once per HEAT stream---regardless of the number of speakers.
}
\label{fig:conditioning}
\vspace{-5mm}
\end{figure}

A natural direction of research has been to adapt high-performing single-speaker ASR architectures to multi-talker scenarios \cite{meng2024empowering, raj2021integration}. Traditional approaches relied on auxiliary speaker embeddings, derived from non-overlapping regions or speaker identification models, to adapt the recognizer to a specific talker \cite{saon2013speaker, kanda2019simutaneous, kanda2022-streamingembedding}. Other systems employed front-end separation modules to disentangle mixed speech signals prior to recognition \cite{kalda2024pixit, lu2021-surt, raj2023-surt2, chen2023speech, yoshioka2018recognizing, sklyar2021-dat, meng2023unified}. More recently, activity-conditioned target-speaker models have gained attention, in which recognition is guided by speaker activity signals obtained from diarization or voice activity detection \cite{polok2024-dicowdiarizationconditionedwhispertarget, polok2024-targetspeakerasrwhisper, wang25-ssa}. These models effectively extend single-speaker backbones to overlapping conditions, but they require the recognizer to be executed once per speaker, limiting practicality for both offline and online tasks. The primary alternative strategy to mitigate this limitation is serialized output training, which reformulates multi-talker ASR as a single sequence generation task, interleaving multiple speaker transcriptions with special tokens to indicate speaker changes \cite{kanda2020serialized}. However, while this approach eliminates the runtime issue, it is sensitive to the annotation format of the segments, requiring hyperparameter tuning to create effective labels \cite{subramanian2025improving}.

In this work, we explore another training strategy to remove speaker count from inference cost in multi-talker ASR while being more robust to annotation styles: Heuristic Error Assignment Training \cite{lu2021-surt, sklyar2021-dat} (HEAT). By discarding speaker attribution and assuming that at most two speakers overlap at any given time \cite{cetin06_interspeech}, HEAT adopts a two-stream formulation where utterances are assigned to non-overlapping, speaker-agnostic streams according to simple heuristics, such as ordering by start time \cite{raj2023-surt2}. These merged streams also have a higher density of speech, which could reduce "leakage" hallucinations when silence-dominant activity masks are used to condition single-speaker ASR models. Although not explored in this work, we note that HEAT references might be easier for front-end systems to produce than diarization labels due to the lack of long-term speaker tracking. 

We also extend HEAT to activity-conditioned target-speaker ASR systems, thereby reducing the number of passes through the recognizer to at most two. A central focus of this study is the design of assignment strategies (i.e. heuristics) that can reliably partition the utterances while maintaining compatibility with pretrained ASR backbones. We validate our approach with Diarization-Conditioned Whisper (DiCoW) \cite{polok2024-dicowdiarizationconditionedwhispertarget, polok2024-targetspeakerasrwhisper}, an existing speaker activity-conditioned target-speaker system, demonstrating that our method substantially reduces computational cost while maintaining competitive recognition accuracy.
To facilitate reproducibility of our experimental setups, we make available our code.\footnote{ https://github.com/xiluohe/heat-conditioned-whisper}

\begin{figure*}[t]
  \centering
  \centerline{\includegraphics[width=\textwidth]{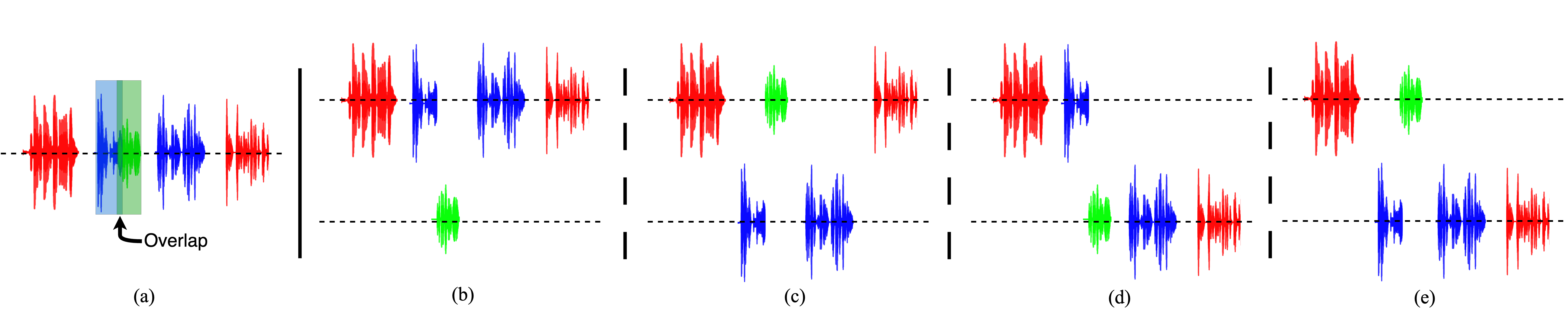}}
\caption{An audio sample (a) with five utterances from three different speakers, as denoted by colors, split into two streams using the following heuristics with HEAT: (b) First-available, (c) Alternating, (d) Recency-continuity, and (e) Speaker-continuity.}
\label{fig:heat}
\vspace{-5mm}
\end{figure*}

\section{Proposed Method}
\label{sec:method}

\subsection{Conditioning ASR Models Using Speaker Activity}

A growing class of target-speaker ASR systems has leveraged speaker activity information, rather than speaker representations, to condition ASR models on a target speaker \cite{polok2024-dicowdiarizationconditionedwhispertarget, polok2024-targetspeakerasrwhisper, wang25-ssa}. Depending on the model, speaker-awareness conditioning is applied before or within the encoder by transforming hidden states $\mathbf{z} \in \mathbb{R}^{d_m \times T}$ with a speaker activity mask $y_{\mathrm{spk}_k} \in [0,1]^T$: $\hat{\mathbf{z}}_t= f(\mathbf{z}_t, y_{spk_k})$.

We extend our work to the Diarization-Conditioned Whisper model (DiCoW) where, rather than conditioning the model solely on binary speaker activity, DiCoW derives a set of four mutually exclusive speaker activity events to guide adaptation \cite{polok2024-targetspeakerasrwhisper}. At each time frame, speech activity is classified into one of four classes: \textbf{S}ilence, only \textbf{T}arget-Speaker, only \textbf{N}on-Target-Speaker, and \textbf{O}verlapping speech between target-speaker and another speaker. The hidden states are transformed using the STNO activity mask $\mathbf{y}_{\mathrm{spk}_k} = [y_{spk_k}^{(S)}, y_{spk_k}^{(T)}, y_{spk_k}^{(N)}, y_{spk_k}^{(O)}]$.

DiCoW implements its conditioning before each transformer encoder layer $l$ by applying one affine transformation for each STNO mask. These four transformed hidden representations are combined through a convex combination weighted by the STNO activity mask called the Frame-Level Diarization Dependent Transformation (FDDT): 
$f(\mathbf{z}_t^l, y_{spk_k}) = \sum_{C\ \in [S,T,N,O] }(\mathbf{W}_C^l \mathbf{z}_t^l + \mathbf{b}_C^l) \odot y_{spk_k}^{(C),t}$.

\subsection{Heuristic Error Assignment Training}
To address the permutation ambiguity problem in multi-talker ASR training, HEAT was proposed as a simplified alternative to Permutation Invariant Training (PIT) \cite{yu2017recognizing}. Whereas PIT computes the training loss across all possible output-target assignments to find the optimal permutation, HEAT uses a deterministic, heuristic-based strategy to assign utterances to fixed output channels to greatly reduce computational cost and memory consumption \cite{sklyar2021-dat, lu2021-surt}. 

Crucial for this work, HEAT is particularly useful in splitting multi-talker audio into non-overlapping streams of voice activity. By assuming at most two overlapping speakers at any given time, HEAT merges utterances into two reference streams such that each stream is non-overlapping \cite{raj2023-surt2}. 
Our work differs from existing HEAT-based systems by extending HEAT to speaker activities masks and through exploring new heuristics.

\subsection{Conditioning ASR Models on HEAT Reference Activities}
In this section, we explain our approach for streaming-amenable multi-talker ASR. The core idea is to convert multi-speaker speech activity into fixed, speaker-agnostic streams using HEAT and then condition the ASR model using these derived activity masks. As shown in \Cref{fig:conditioning}, given speaker activity masks \(\mathbf{y}_{\text{spk}} \in [0,1]^{T \times K}\), obtained from an external diarization system, we extract a set of \(N\) utterance-level segments \(U = \{ u_1, \dots, u_N \}\) by identifying all contiguous stretches of activity for every speaker speaker before merging them into two streams \(\mathbf{y}_{\text{HEAT}} \in [0,1]^{T \times 2}\). The resulting speaker-agnostic activities is used to derive the two streams' supervision and conditioning mask. Although speaker identity is not preserved in this process, the model can be trained efficiently by computing the loss only once per stream and reduces runtime inference by limiting the number of encoder forward passes required (ie. reduce the number of streams to process). Additionally, using speaker-agnostic activity masks removes the need for long-term speaker tracking, eliminating the speaker clustering step of speaker-activity extraction that is not amenable to streaming.

Integrating HEAT into existing conversational target-speaker ASR models requires a stream assignment heuristic that satisfies several properties critical for both training stability and recognition accuracy. Specifically, we want a heuristic that has (1) runtime and performance consistent with speaker-based systems for single- and two-speaker scenarios, (2) balanced speech activities between both streams to prevent dominance bias by one stream, and (3) stream-wise continuity so utterances belonging to the same non-overlapping local dialogue region are not split up. 

Balanced speech activities is important because, as explained in \cite{wang25-ssa}, single-speaker ASR systems tend to prioritize a single speaker while disregarding the other speakers due to the encoder states being optimized for a particular speaker's accuracy. We found that conditioning the model heavily imbalanced activity streams leads to insubstantial weight adjustments that cannot counteract this bias. Additionally, preserving continuity is important for the language model to use local context for more accurate decoding, as well as mitigating the effect of diarization errors that segment continuous speech \cite{linke2025context, masumura2023end}.

To construct references, we first define a stream being "available" for an utterance if, for the entire duration of that utterance, that stream does not contain any other speech activity. The simplest way to construct HEAT streams comes from \cite{raj2023-surt2}, which uses the \textbf{first-available} heuristic. First, the utterances are ordered by start times and sequentially assigned to a stream. Then, for each utterance, if the first stream is available, it is assigned to the first stream. Otherwise, it is assigned to the second stream. While overlapping speech are separated, this does not satisfy most of our criteria. 

As shown in \Cref{fig:heat}, we propose the following three heuristics for conversational settings that focuses on picking a stream when both streams are available. These heuristics also assign utterances sequentially and will first check if both streams are available. If only one stream is available, then that stream is chosen. If both streams are available, then (1) the \textbf{alternating} heuristic assigns utterances to the stream opposite the previous utterances' assignment.
(2) The \textbf{recency-continuity} heuristic assigns utterances to the most recently active stream. (3) If either stream left off with the current utterance's speaker, the \textbf{speaker-continuity} heuristic assigns the utterance to that stream; if not, fallback to the \textit{recency-continuity} heuristic.
 


\section{Data and Experimental Setup}
\label{sec:experiments}

\subsection{Data}
We train our models on the AMI corpus and evaluate on AMI, ICSI, and LibriMix \cite{kraaij2005ami, janin2003icsi, cosentino2020librimix}. For AMI and ICSI, we use the single distant microphone (SDM) condition. AMI contains approximately 100 hours of multi-speaker meetings involving four to five participants per session and exhibits a two-speaker overlap rate of 22.1\% in the training set and 21.0\% in the test set, with more than two speakers active in 3.4\% and 6.0\% of frames, respectively. ICSI comprises around 72 hours of conversational speech recorded in real meetings involving three to ten speakers. The two-speaker overlap in ICSI accounts for 9.0\% of training data and 13.6\% of test data, while more than two speakers are active only rarely (0.7\% and 1.4\% of frames). To evaluate performance under controlled overlap conditions, we also use the sparsely overlapping version of LibriMix (SparseLibriMix), which mixes Librispeech utterances to achieve varying 2- and 3-speaker overlap ratios.

\begin{table}[t]
\centering
\caption{Comparison of different HEAT heuristics derived from oracle speaker activities to condition Whisper. All reported values are tcORC-WER.}
\label{table:results-heuristics}
\vspace{0.2cm}
\begin{tabular}{@{}lcc@{}}
\toprule
Heuristic                  & AMI-SDM ($\downarrow$) & ICSI-SDM ($\downarrow$) \\ \midrule
First-available                               & 32.41           & 40.45                \\
Alternating                            & 22.20           & 25.47                \\
Recency-continuity                & 20.64           & 24.42               \\
Speaker-continuity  & 19.71               & 24.94                    \\
\midrule
Diarization                                  & 17.18           & 23.84                \\ \bottomrule
\end{tabular}
\vspace{-5mm}
\end{table}

\subsection{Training Details}
To compare our approach with an existing speaker activity conditioned ASR model, we use DiCoW as our baseline. To integrate HEAT streams into DiCoW, we condition Whisper with similarly created STNO masks but with target streams instead of speakers. In line with DiCoW, we use \textit{Whisper-large-v3-turbo} with an additional Connectionist Temporal Classification (CTC) head and two convolutional layers, both with a subsampling factor of two \cite{polok2024-targetspeakerasrwhisper}. The CTC head and the decoder are both trained with timestamp tokens, and the CTC loss weight is fixed at $0.3$. 

All models are trained for 10 epochs with an adaptive batch size using the AdamW optimizer. The base learning rate is $2 \times 10^{-6}$, with a weight decay of $1 \times 10^{-6}$, a linear decay schedule, and $2000$ warm-up steps. Parameters introduced by FDDT are trained with a learning rate of $2 \times 10^{-4}$. 

For inference, we use greedy decoding unless beam decoding is specified. Beam decoding is performed with 5 beams, a length penalty of $0.1$, and a CTC weight of $0.2$.  

\subsection{Evaluation}
Since we do not have speaker labels, we measure ASR performance through time-constrained optimal reference combination word error rate \textit{tcORC-WER}. ORC-WER calculates speaker-agnostic multi-talker word error rate by finding the optimal assignment of reference utterances across output streams while preserving speakers' temporal order. 
We use the time constrained version of this metric with a five second collar since hypotheses and references can be aligned across implausibly long temporal distances \cite{vonneumann2023-orcwer}.

To measure relative runtime, we also calculate inverse real time factor (RTFx). RTFx finds the length of audio that the system can process (ie. preprocess, encode, decode, and postprocess) in one second:
$\text{RTFx} = \frac{\sum_{i=1}^{N} T^{(i)}_{\text{audio}}}{\sum_{i=1}^{N} T^{(i)}_{\text{proc}}}$.

\begin{table}[t]
\centering
\caption{\textbf{Inference efficiency with pretrained diarization model outputs.} Comparison of model inference times when using conditioning Whisper with speaker activity masks derived from Diarizen and decoding with beam search (n=5). The HEAT masks are generated with the speaker-continuity heuristic. Reported WER corresponds to tcORC-WER.}

\label{table:results-timefactor}
\vspace{0.2cm}
\begin{tabular}{@{}lcccc@{}}
\toprule
 & \multicolumn{2}{c}{AMI-SDM} & \multicolumn{2}{c}{ICSI-SDM} \\ \cmidrule(l){2-5} 
Activity Mask        & WER($\downarrow$)      & RTFx($\uparrow$)      & WER($\downarrow$)       & RTFx($\uparrow$)       \\ \midrule
Speaker  & 18.34          & 2.05      & 25.55               & 1.50          \\
HEAT         & 18.99          & 4.57      & 26.24               & 3.89          \\ \bottomrule
\end{tabular}
\vspace{-5mm}
\end{table}

\section{Results and Discussion}

\Cref{table:results-heuristics} compares the proposed method under different HEAT stream-assignment heuristics, using oracle speaker activities for conditioning. It can be seen that naively merging streams with the \textit{first-available} heuristic leads to model collapse, where both streams converge to nearly identical transcripts, causing high tcORC-WERs. Additionally, comparing the \textit{alternating} and \textit{recency-continuity} assignment strategies demonstrates the importance of preserving local conversational continuity as \textit{alternating} retain very little continuity. The best-performing strategy, \textit{speaker-continuity}, approaches the accuracy of directly using speaker activities. Its effectiveness can be attributed to balancing speech activity more evenly across streams while maintaining local speaker and context continuity. On ICSI, due to the large number of speakers, the \textit{speaker-continuity} strategy often falls back to the \textit{recency-continuity} case, leading to similar performance for the two heuristics. Due to the strong performance of \textit{speaker-continuity}, we use it for the rest of our experiments.

\Cref{table:results-timefactor} reports results using outputs from an automatic diarization system rather than oracle annotations. For this experiment, we employ Diarizen \cite{han2025diarizen}, a pretrained end-to-end diarization model trained on meeting data, to obtain speaker activity masks that are merged into HEAT streams. Compared to the oracle case, the performance gap between diarization-based conditioning and HEAT-based conditioning is reduced, since the diarization model itself struggles to accurately resolve overlapping speech segments. In terms of efficiency, HEAT achieves a 123\% relative improvement in RTFx on AMI and a 159\% relative improvement on ICSI. The larger gains on ICSI can be attributed to its higher average number of speakers per recording, which amplifies the runtime cost of diarization conditioning. Finally, we note that the diarization model over-predicted the number of speakers for some samples in our test set, further increasing runtimes. 

\Cref{table:results-sparselibrimix} presents a comparison of diarization-based conditioning and HEAT-based conditioning under varying levels of two- and three- speaker overlap on the synthetic SparseLibriMix test set. As expected, the introduction of three-speaker overlap, even small amounts, widens the gap between HEAT conditioning and diarization conditioning, reflecting the limitations of the two-speaker overlap assumption. 

\begin{table}[t]
\centering
\caption{{\textbf{Impact of Overlapping Speech.} tcORC-WER of different overlap conditions on the SparseLibriMix dataset. Overlap percentages represent the proportion of 2- or 3-speaker overlapping speaking time.}}
\label{table:results-sparselibrimix}
\vspace{0.2cm}
\begin{tabular}{@{}lcccc@{}}
\toprule
      & \multicolumn{2}{c}{2-Speaker} & \multicolumn{2}{c}{3-Speaker} \\ \cmidrule(l){2-5} 
\% OV & Speaker($\downarrow$)       & HEAT($\downarrow$)       & Speaker($\downarrow$)      & HEAT($\downarrow$)       \\ \midrule
0     & 6.56             & 6.34       & 7.23            & 6.93       \\
5     & 7.13             & 6.48       & 24.02            & 26.22      \\
10    & 7.90             & 7.48       & 23.48            & 26.16      \\
15    & 9.00             & 8.40       & 23.97            & 26.97      \\
20    & 10.04            & 11.83      & 30.29            & 33.10      \\ \bottomrule
\end{tabular}
\vspace{-5mm}
\end{table}

\section{Conclusion}
In this study, we introduced a HEAT-based activity conditioning framework to decouple inference cost from the number of active speakers, focusing on effective speaker activity merging heuristics. We demonstrated its effectiveness by integrating it into DiCoW using both oracle and model-extracted speaker activities.

While our proposed HEAT heuristics proves effective under DiCoW, further validation with alternative model architectures (particularly streaming backbones) or new activity conditioning schemes is needed.
Future work also includes adding speaker attribution back to the output and training a system to directly output HEAT streams, instead of constructing them out of diarization model outputs, as it enables end-to-end training of the multi-talker ASR system.

\bibliographystyle{IEEEbib_abbrev}{\footnotesize
\bibliography{refs}}

\end{document}